\documentstyle[12pt]{article}
\title{\bf Extragalactic Objects and Next Generation Interferometers 
}
\author{D.~Fraix-Burnet %\thanks{This is a footnote in the author list},
\\
\vspace{1cm}\\
\normalsize Laboratoire d'Astrophysique de Grenoble, France}
\date{\mbox{}}
\begin{document}
\maketitle
\pagestyle{empty}
%
% WE REDEFINE THE plain LaTeX PAGESTYLE !!! 
% THIS PAGESTYLE WILL BE USED FOR THE FIRST PAGE ONLY !
%
\def\bull{\vrule height .9ex width .8ex depth -.1ex}
\makeatletter
\def\ps@plain{\let\@mkboth\gobbletwo
\def\@oddhead{}\def\@oddfoot{\hfil\tiny\bull\quad
``Science Case for Next Generation Optical/Infrared Interferometric Facilities 
(the post VLTI era)'';
37$^{\mbox{\rm th}}$ Li\`ege\ Int.\ Astroph.\ Coll., 2004\quad\bull}%
\def\@evenhead{}\let\@evenfoot\@oddfoot}
\makeatother
%
% AND DEFINE OUR MACROS FOR THE REFERENCE LIST
% I.E \beginrefer \refer and \endrefer
%
\def\beginrefer{\section*{References}%
\begin{quotation}\mbox{}\par}
\def\refer#1\par{{\setlength{\parindent}{-\leftmargin}\indent#1\par}}
\def\endrefer{\end{quotation}}
%
% BEGIN THE ABSTRACT CHAPTER WITH \noindent\small, ENCLOSE IT IN A GROUP
% AND BOLDFACE THE TITLE.
%
{\noindent\small{\bf Abstract:} 
 The most obvious extragalactic targets for optical/infrared interferometers are Active Galactic Nuclei. In this document, I try to overview other topics that could be of interest to studies of galaxies and whether they could be adequate for a next generation interferometer. The very high spatial resolution would be profitable for extragalactic supernovae, globular clusters, star forming regions, gravitational lenses and some stellar studies in very close galaxies. However, sensitivity is the main concern since the interesting magnitude limit would have to be of the order of 25 or more for these studies not to remain marginal. 
 }
%
% NOW COMES THE MAIN BODY OF THE ARTICLE
%

\section{Introduction}

Galaxies are independent groups of stars, gas and dust. These constituents evolve both in time and in space, in the sense that galaxy evolution depends much on the environment and thus does not follow the same tracks in different places in the Universe. The principal question of extragalactic astronomy at the beginning of the XXIst century is to understand galaxy formation and evolution. Are galaxies formed once with a large ``initial mass function'' (monolithic scenario) or are they all formed by merging and accretion of small building blocks (like dwarf galaxies) which could have been the first objects of the Universe?

Traditionally, galaxies were studied globally as whole entities because they could not be resolved in individual stars. Overall shapes and some conspicuous features were determined from imaging. High and medium spectroscopy provided  dynamics and chemical composition from which galaxy ages were estimated. This ``global'' approach is still used for distant galaxies in which evolution is so clearly observed. 

Going to high redshifts is not sufficient to understand galaxy evolution. 
Improved spatial resolution and progresses in the knowledge of our own Galaxy have revealed the necessity to consider galaxies as mixtures of very different stellar populations, gas regions and dust clouds. It is then now obvious that the understanding of galaxy evolution asks for the detailed observations of all these ingredients. After our own Galaxy, this of course can only be done in nearby galaxies. The detailed physics of these ingredients, particularly the stars, do rely on observations within our Galaxy. But the distance determinations are not easy, less than in another galaxy for which the distance of all the objects is the same. In addition, the physics of the objects depends on the environment such as the local metallicity, and only extragalactic studies can provide an idea on the universality of our knowledges. Optical/infrared interferometers provide very high spatial resolutions and thus are well suited to map small and/or distant objects and to individualize objects in crowded fields (the so-called confusion problem).

Sensitivity is obviously a key point in extragalactic observations and should be the next important step to be conquered by interferometry. Field of view can be often a serious limitation. Another point of importance is the lack of bright point-like reference source for adaptive optics and fringe tracking close to extragalactic objects. This could however be solved thanks to technological progress, efficient use of artificial/laser stars and increase of the field of view. Imaging capabilities are absolutely required when dealing with complex objects like galaxies. ELTs would perform much better in mapping entire galaxies because of their much higher sensitivity and their larger field of view. Nevertheless, interferometers must have a reasonable number of apertures to map at once individual variable objects like supernovae remnants, and true imaging capabilities would be much appreciated in studying star forming regions in close galaxies. It is also not conceivable to imagine looking at other galaxies star by star: a lot of objects should be present in the field of view and analysed at the same time. In addition, except for some stars, astronomical objects are 2-D when projected on the plane of the sky and not centrosymmetric. Models are thus 2-D at least and should be constrained in the uv-plane, not on the visibility curves.

In this document, we exclude the central parts like Active Galactic Nuclei that are reviewed elsewhere in this conference (Marconi). The instrumental environment is set by current performances of optical/infrared interferometers (expected sensitivity up to K=20 for AMBER/VLTI with three 8m telescope and a baseline of 80 meters) and envisaged ELTs (expected sensitivity up to the 35-38 for a diameter of 100 meters - 1~mas resolution). Logically, the next generation of interferometers should be more sensitive than current ones and reach higher resolutions than ELTs.

In a first section (Sect~\ref{goals}) I overview several a priori possible interesting science goals suited for an interferometer, without aiming at a complete review for each. I try to summarize the technical requirements in Sect~\ref{specs}.

\section{Science goals}
\label{goals}

There are two main categories of galaxy observations. The first one considers galaxies globally, without resolving them into individual stars or clusters. The first three goals below belong to this category, and require both good imaging capabilities and excellent sensitivity, except may be for the multiple images of quasars due to gravitational lensing. The second category of observations consider individual objects within galaxies. Because of sensitivity requirements, they are limited to relatively close objects, and the large number of targets probably imply a significantly large field of view.

\subsection{Galaxy imaging}

The goal is to study in detail the nature of galaxies glimpsed for instance in the Hubble Deep Field or other surveys of distant galaxies. Their shapes appear to be perturbed at high redshifts that could be explained by a period of high interaction/merger rate (Abraham \& van den Bergh 2001). Imaging at very high spatial resolution (1~mas or better) could try to identify and characterize spiral arms, bars, merger signatures, tails, multiple cores, dust distribution, large star forming regions. Spectrophotometric data could also give their global kinematics and metallicity. Because the description of galaxies based on their global morphology like in the Hubble diagram is now recognized to be not sufficient (Abraham \& van den Bergh 2001), all these data would be invaluable to understand the evolutionary history of galaxies and their constituents (stars, gas and dust). These very distant objects put some more and more severe constraints on galaxy formation scenarios because structures and composition impose a time scale since the very first objects in the Universe. However, ELTs can probably do better in this domain due to their better imaging capability, sensitivity and field of view. Unfortunately, the next generation of interferometers will certainly not be able to bring any complementary observation, and imaging of closer galaxies necessitates a field of view that could never be reached by interferometers.

\subsection{Galaxy 2D spectroscopy: kinematics, metallicity, chemical composition}

The medium and high spectroscopy of galaxies contain all the interesting information: chemical and kinematical, the last one describing the galaxy morphology. 
High spatial resolution is also important to study separately the different components of a galaxy: the bulge, the disk and the starburst. Each of them has a different origin and history which is not quite known. They all trace different evolutionary phenomena, on different time scales, and probably exchange material with one another. In this purpose, 
medium spectral resolution would give strong indications on the evolution of these components as well as the other structures (like bars) and the different stellar populations within a galaxy. High spectral resolution is necessary to precise the metallicity and the chemical composition of stellar groups. However, a very high spatial resolution might not be that crucial here, especially if there are strong sensitivity limitations and small field of views. Hence,
interferometers are not yet the best instruments for this kind of study.

\subsection{Gravitational lensing}

The distant Universe is full of distorted images of galaxies (Abraham \& van den Bergh 2001). These are the gravitational lenses that bring two kinds of opportunities. Firstly, by amplification of the light, they provide a good view of very distant and faint objects that would otherwise be undetectable. Provided the mass distribution of the lens is known, it is possible to
reconstruct the precise galaxy morphology by using some details within the distorted image. Secondly, gravitational lenses trace the mass distribution of lensing galaxies and clusters, including the dark matter. A precise cartography of many distorted images for the same lens is necessary. This could be a good target for interferometers, but unfortunately the surface brightness is far too low for next generation instruments. 

Interferometers can however be useful in the study of multiple images of quasars due to gravitational lenses. About a hundred of such objects are known (Kochanek et al 2004) with \textit{I} magnitudes ranging between 14 and 24. This number depends much on the resolution that is currently at best a few tenths of an arcsec. With much higher resolution much more lenses could be detected and studied with better details. 
Monitoring any variable event and measuring the time delay between the different images of the same quasar gives an interesting measure of the cosmological parameters (Blandford \& Narayan 1992). For this purpose, a very good astrometry of the different images is required (Witt et al 2000). Detailed analysis of the lensed system also provides indications on the mass distribution in the lensing galaxy. A good sensitivity is necessary but probably within the range of a next generation interferometer. 

\medskip

All these global studies of galaxies presented in the three previous sections are thus better targets for ELTs until very large arrays with much improved sensitivity and larger field of view can be designed. The only exception is gravitational lenses with multiple images of quasars that seem to be within the reach of a next generation instrument providing it has some imaging capabilities and a sensitivity reaching the 24th magnitudes at least.

\subsection{Individual stars: population, IMF}

Our knowledge of stellar population and history is largely based on detailed studies within our own Galaxy. How universal are our schemes? This will find answers by repeating the same kind of observations in other galaxies. Galaxy evolution itself is mainly described by the evolution of its stellar population. We know that a galaxy is a mixture resulting from several stellar formation events. Identifying the different ``generations'' would be essential to understand the galaxy history and more generally the stellar formation history in the Universe. Since this requires high spatial resolution to individualize stars as much as possible, it has been done in our own Galaxy and some other very close Local Group members. Going farther away necessitates higher spatial resolution and sensitivity.

Colour-magnitude diagrams coupled with metallicity measurements of individual stars in other galaxies will precisely trace the star formation history by determining the age and chemical composition of groups of stars. Their kinematics also tells much about their origin (accretion, galaxy merger, in situ formation). The study of specific stars (Be, LL Lyrae, Brown Dwarfs, etc...) in other galaxies will add more insights in the detailed physics of stars. 

In this kind of studies, the two main problems for interferometers are field of view and sensitivity. It is unrealistic to imagine observing a significant fraction of the stellar population of a galaxy one star after one star! A relatively large field of view is required. ELTs will do much better in this respect, except in very crowded regions were a very high resolution would help individualize stars. The other problem is sensitivity: observing a star in a galaxy of the Virgo cluster is certainly a challenge for optical interferometers (see Table~\ref{numbers}), even though it would be important because it is where the closest elliptical galaxies are located. Nevertheless, one could hope to identify and to study groups of stars, which are brighter than individual stars, belonging to several regions with different stellar populations. 

In somewhat closer galaxies, detailed observations of star formation regions would probably be the best way to establish the Initial Mass Function (IMF) while assessing its universality. It is a crucial ingredient for star formation and galaxy evolution models. In addition, direct mass measurements from double stars could be possible in nearby dwarf galaxies with a resolution better than 0.01~mas. The observations should not be limited to the brightest stars, hence such studies would be more complete with an interferometer reaching a magnitude limit of 25.

\subsection{Globular clusters}

The determination of the age of the stellar population of globular clusters requires resolution of individual stars using photometry and high resolution spectroscopy. The formation history of globular clusters can thus be constrained and compared with the properties of their parent galaxies: do they form during merger events? Do they form in situ within the galaxy? Are they associated with starbursts events? When and how do they form? What is the difference between globular clusters and dwarf galaxies? Separating stars individually is clearly in favour of interferometers as compared to ELTs, but sensitivity would impose the same limit as other stellar studies, that is mainly for galaxies of our Local Group.

\subsection{Star forming regions, starbursts, HII regions, planetary nebulae}

As already mentioned, study of individual stars might be limited to relatively close galaxies because of sensitivity. But integrated magnitudes of stellar formation, starburst and HII regions can certainly be identified in Virgo cluster galaxies at least. Spectroscopy could also be possible and investigation of the way stars form in different galaxies will bring fundamental knowledge about galaxy evolution. Extreme star forming galaxies that are Ultraluminous Infrared Galaxies like Arp220 (at 75 Mpc) must be observed with very high spatial resolution to map the distribution of star forming regions. The long standing question about the power source (AGN versus starburst) will certainly find an answer, especially if the sensitivity is sufficient to observe galaxies far away enough.
However, the advantage of an interferometer over a 100m ELT might not be that enormous except in somewhat crowded cases such as starburst irregular galaxies like M82. But rapidly the surface brightness of these objects will impose a severe limitation on the number of reachable targets for the next generation interferometers.

\subsection{The interstellar medium}

The very high spatial resolution could allow a mapping of the interstellar medium in its different phases and the dust clouds in other galaxies as distant as possible. The nature of dust grains can be constrained by observations in the near-IR while the physical conditions of the gas is obtained in the optical. This kind of study clearly involves full imaging capability with an excellent surface brightness sensitivity associated with a reasonable field of view. In some cases, one might be interested in studying some absorption characteristics along a very thin beam. But globally the usefulness of an interferometer to study the interstellar medium would be probably marginal.

\subsection{Supernovae}

Supernovae are an important ingredient in the physics of stars, their IMF and the way they interact with their environment. In our Galaxy, SNRs are observed quite in detail. But the sample is limited and affected by great distance uncertainties. Looking into another galaxy alleviates these two difficulties, and multiplying the number of galaxies would still increase the sample. SNRs have already been observed or detected in other galaxies like M31. However, only in the LMC a good vision and follow-up of the ejecta is possible. Observing supernovae in details in several galaxies will show how they behave in other environments, and more importantly will increase the number of supernova explosion analyses that would be so important to better understand their physics (e.g. Weiler \& Sramek 1988, Chen et al 2002). 
This is fundamental because supernovae can also serve as cosmological candles. This constitutes one of the methods to determine the Hubble constant (Mould et al 2000).
The sensitivity  of an interferometer would probably limit observations to young supernova remnants, when they are bright and not too extended and diluted, at least as far as the Andromeda galaxy and maybe a little bit farther away. But this is exactly soon after the explosion that spatial resolution is the most needed! It will allow us to follow the expansion of the ejecta like in the spectacular case of SN1987A but much sooner after the explosion and even before an ELT can resolve the expanding shell. This is certainly a good target for interferometers even if the number of supernova explosion in the close Universe is not very frequent.

\subsection{Gamma Ray Bursts}

GRBs are now known to be at cosmological distances, and the current belief is that they are the result of the collapse of a massive star, maybe somewhat like an hypernova (e.g. Chen et al 2002, Garvanich et al 2003, Price et al 2003). Because they are extremely bright, they are observed at extremely high redshifts, but the parent galaxies is consequently extremely difficult to detect. However, it would be essential to know better in which kind of objects they are located, and where. Here, sensitivity is crucial as well probably as a good contrast capability. Like for supernovae, with vey high spatial resolution, it could be possible to follow the expansion of the explosion. However, the GRBs star fades rapidly, in a few weeks, beyond magnitude 25 so that it seems rather hopeless for a next generation interferometer to observe the expanding shell at such cosmological distances. However, it could probably help in determining very accurately the location of the GRB event with respect to the parent galaxy that could be imaged by a ELT (Berger et al 2003, Courty et al 2004). It could also follow precisely the afterglow brightness curve by avoiding source confusion over a few days.

\subsection{Cepheid and the distance scale calibration}

Cepheid stars are the primary distance indicator on which all others are based. In particular, the recent HST Hubble constant measurement utilises SNIa type supernovae (Freedman et al 2001) and depends entirely on the calibration using Cepheids. Determination of the zero point of the Period-Luminosity relation is thus crucial. This can be done in our own Galaxy and the extension to other Local Group galaxies and even farther ones considerably increases the accuracy of the calibration (Fouqu\'e et al 2003). But this requires an independent and accurate knowledge of their distance. Interferometers, even current ones for our Galaxy, can alleviate this difficulty by measuring simultaneously the angular diameter and the radial velocity amplitude variations to derive the linear diameter (Lane et al 2002, Kervella et al 2004a, Kervella et al 2004b). Because Cepheids are massive stars, they are bright enough to be observed by an interferometer in the Virgo cluster but measuring their angular diameter cannot be measured farther away than the Local Group.

\subsection{Conclusion on possible targets}

As a summary, only a very few extragalactic topics can be covered by a next generation interferometer apart from AGNs. These are gravitational lenses (multiple quasar images), stellar studies in very close galaxies, star forming regions, globular clusters and supernovae remnants.

\section{Instrumental requirements}
\label{specs}

\subsection{Distances and sizes: angular resolution}

The main power of interferometry is its high angular resolution. Present instruments have baselines of the order of 100 meters which is also the maximum size of foreseen ELTs. This give an angular resolution of about 1~mas at 1~micron wavelength. Obviously, next generation interferometers should do much better. But what kind of baselines would extragalactic targets require? To help answer the question, Table~\ref{numbers} give some typical sizes of some objects and the corresponding angular size at different distances: 200~kpc which is about the distance to the nearest dwarf galaxies of the Local Group, 1~Mpc which is the distance between the Andromeda galaxy M31 (750~kpc) and the closest starburst galaxy M82 (3~Mpc), 16~Mpc which is the distance to the Virgo cluster.

Measuring Cepheid diameters in close dwarf galaxies would require a 100~km baseline at least, while 10~km is sufficient to resolve some double stars. Stellar crowdedness much depends on the region of a galaxy, but the gain over a 100~m ELT can be quite large even with a 1~km baseline. It should be noticed that of course stellar distribution is 3-D so that in very dense regions, foreground and background stars will limit the advantage of an interferometer. However, a 10~km baseline would be of great help in individualizing stars in dense regions such as globular clusters or central parts of galaxies. 

Stellar groups (globular clusters and star forming regions) have sizes easily resolved by an interferometer in the Virgo cluster and even beyond. However, the goal is to individualize stars, and in relatively close objects the problem becomes the field of view. In these objects, both a resolution of about 0.1~mas \emph{and} a field of view between a few hundreds and 1000~mas seem to be required. Mosaicing could be a solution to compensate for the latter.

Gravitational lenses and supernova remnants are not mentioned here because anything much better than 1~mas would be  sufficient. Yet, some imaging capability is necessary. This implies a range of baselines to be covered from 100~m and above to ensure full complementarity and ease of comparison with other instruments, ELTs in particular. 

As a conclusion, it seems that an array with baselines between 100~m and about 10~km is well suited for extragalactic observations. 

\begin{table}
\begin{center}
	\begin{tabular}{|l|c|c|c|c|c|c|c|}
\hline	\hline
	Object              & Typical      & 	\multicolumn{2}{c|}{200 kpc}&\multicolumn{2}{c|}{1 Mpc}     & \multicolumn{2}{c|}{16 Mpc}    \\
	                    &  size        & \multicolumn{2}{c|}{(close Dwarfs)} & \multicolumn{2}{c|}{(Andromeda)}  & \multicolumn{2}{c|}{(Virgo)}   \\
	                    \hline
	                    &              &  \textit{mas} & \textit{mag} &  \textit{mas} & \textit{mag} &  \textit{mas} & \textit{mag} \\
	\hline\hline
Cepheid               &60 R$_{\odot}$& 0.001 &18&-&-&-&-   \\
	\hline
Double star           &     10 AU    & 0.01  &20&0.002&23&-&-   \\
	\hline
Stellar crowdedness   &   0.1 pc     & 20    &20&4&23&0.2& 30 \\
	\hline
	Globular clusters   &   5 pc       & 1000  &12&200&15&70&22   \\
	\hline
  Compact HII region  &   50 pc      & 10000 &15&2000&18&700&25  \\
\hline \hline
	\end{tabular}
\end{center}
\caption{Angular sizes and magnitudes of some typical objects found in galaxies, at different distances.The size given for ``Double star'' and ``Stellar crowdness'' corresponds to the separation between stars, whereas the magnitude is the magnitude of individual stars. For the globular cluster and compact HII regions, the size and magnitude are integrated values for the whole system.}
\label{numbers}
\end{table}

\subsection{Intensities: sensitivity}

Sensitivity is both a serious limitation of interferometers and a key point for extensive extragalactic observations. 
A typical galaxy has an integrated magnitude of the order of 26 to 30 in R at a redshift of 4. A supernova like SNIa would have I$\simeq 29$ at this distance but the remnant itself is fainter. In the first case, the surface brightness is definitively too low for interferometers, whereas in the second case observations should be possible in relatively close galaxies. A GRB is at best K$\simeq 18-20$ at z=10 soon right after the burst and then dim quickly (days to weeks). Except for pinpointing precisely the position of the exploding source, an interferometer is not suited for these kind of objects because either the ejecta is far too small to be resolved even with a very long baseline or it is too faint. 

Table~\ref{numbers} gives the typical apparent magnitudes of some star related objects at different distances from the observer. The average apparent magnitude of a star in a Virgo cluster galaxy is about 30. Stars of the main sequence are fainter and would have an apparent magnitude of the order of 32 at this distance and about 25 in Local Group galaxies. Evolved low-mass stars (RGB, AGB) and massive stars are brighter but do not represent the entire stellar population. 
Globular clusters and star forming regions have integrated apparent magnitudes below 25 in the Virgo cluster but resolving them into individual stars would require a much higher sensitivity. 

As a summary, an extensive extragalactic program would require reaching magnitudes well above 30 and this is the goal of ELTs. Nevertheless, forgetting objects at cosmological distances, it is possible to do very broad and interesting studies of galaxies with a lower sensitivity. Naturally, it should not be too low: the necessary limiting magnitude for a next generation interferometer seems to be around 25. 

\subsection{Summary of specifications for extragalactic observations}

The requirements in term of spatial resolution and sensitivity mentioned in the two previous sections are translated into instrumental specifications in Table~\ref{requirements}.
All identified topics would be satisfied at both optical and near infrared wavelengths. High spectral resolution for chemical composition and metallicity means R=5000-10000, but medium spectral resolution is sufficient for galaxy kinematics. No polarization is necessary for the moment. Fringe tracking is compulsory to allow for long exposure times but since the probability of finding a relatively bright reference source very close to the target is thin, off-axis fringe tracking and/or artificial stars should be implemented. Imaging capability, at least in the sense of model fitting of radio astronomers in a more or less dense uv-plane, is an absolute requisite for extragalactic observations.

\begin{table}
	\begin{center}
		\begin{tabular}{|l|l|}
		\hline
		\hline
			Sensitivity &  mag$\geq 25$ \\
\hline
			Observing mode & Direct imaging\\
\hline
			Spatial resolution & $<0.1$ to 1~mas\\
\hline
			uv-coverage & As dense as possible, with \\
			            & small (100~m) and long (1-10~km) baselines \\
\hline
			Wavelength range  & optical - NIR \\
\hline
			Spectral resolution & 0-10000 \\
\hline
			Polarimetry & None \\
\hline
			Field of view & as large as possible \\
\hline
			Contrast & 100-1000 \\
\hline
			Astrometric precision & Standard \\
\hline
			Maximum observation time & Long exposures (hrs) \\
\hline
			Others & Off-axis fringe tracking \\
			       & Artificial/laser star useful \\
			       & Phase closure \\
		\hline
		\hline
		\end{tabular}
	\end{center}

	\caption{Summary of instrumental requirements for a next generation interferometer in the case of extragalactic studies apart from AGNs.}
	\label{requirements}
\end{table}

\section{Conclusion}

The richness of extragalactic studies much depends on the level of sensitivity an interferometer can reach. Ideally, magnitudes slightly above 30 should be attained, so that a very important fraction of the Universe can be observe. This is about 10 magnitudes more than what current interferometers are supposed to do. If this ideal limit cannot be reached, then all ``global'' studies of galaxies should be left to ELTs. But still, the 25th magnitude is necessary for stellar studies in other galaxies.

Regarding spatial resolution, there is not much constraint on the optimum baseline because there are so many different scales and distances. However, because of this, the next generation interferometers must have many apertures, about ten or even more, spanning a whole range of baselines, in order to constrain 2-D models of targets that are nearly always quite complex. This should certainly be associated with a field of view as large as possible in order to include several stellar objects in one and same observation.

In these conditions, interferometers can bring crucial information to improve our understanding of stellar physics and galaxy evolution as deduced from very detailed studies of close galaxies. ELTs will probably concentrate on faint, large, complex and distant objects, while interferometers will be used for stellar studies in crowded environments in close galaxies, star forming regions, globular clusters and supernovae remnants. The only possible cosmological targets for interferometers seem to be gravitational lenses of multiple quasar images for which excellent spatial resolution and astrometry accuracy are required.

%
% USE A SECTION WITHOUT NUMBER FOR THE ACKNOWLEDGEMENTS
%
%\section*{Acknowledgements}
%This research was supported in part by contract ARC 90/94-140 ``Action de
%Recherche Concert\'ee de la Communaut\'e Fran\c{c}aise'' (Belgium).
%
% BEGIN THE REFERENCE LIST WITH \beginrefer
% USE \refer BEFORE THE REFERENCES AND BEGIN A NEW PARAGRAPH AFTER THE 
% REFERENCE !
% DO NOT FORGET TO END THE LIST WITH \endrefer
%
 
\beginrefer

\refer Abraham R. G., van den Bergh S., 2001, Science, 293, 1273

\refer Blandford R.D., Narayan R., 1992, ARAA, 30, 311

\refer Berger E., Cowie L.L., Kulkarni S.R., Frail D.A., Aussel H., Barger A.J., 2003, ApJ, 588, 99 (astro-ph/0210645)

\refer Chen C.-H.R., Chu Y.-H., Gruendl R., Lai S.-P., Wang Q.D., 2002, AJ, 123, 2462 (astro-ph/0202047)

\refer Courty S., Björnsson G., Gudmundsson E.H., 2004, MNRAS, in press (astro-ph/0407359)

\refer Fouqu\'e P., Storm J., Gieren W., 2003, Lect. Notes Phys., 635, 21 (astro-ph/0301291)

\refer Freedman et al, 2001, ApJ, 553, 47 (astro-ph/0012376)

\refer Garnavich P.M., Stanek K.Z., Wyrzykowski L., Infante L., Bendek E., Bersier D., Holland S.T., Jha S., Matheson T., Kirshner R.P., Krisciunas K., Phillips M.M., Carlberg R.G., 2003, ApJ., 582, 924 (astro-ph/0204234)

\refer Kervella P., Nardetto N., Bersier D., Mourard D., Coud\'e du Foresto V., 2004a, A\&A, 416, 941

\refer Kervella P., Bersier D., Mourard D., Nardetto N., Coud\'e du Foresto V., 2004b, A\&A, in press (astro-ph/0404179)

\refer Kochanek C.S., Falco E.E., Impey C., Lehar J., McLeod B., Rix H.-W., 2004, http://cfa-www.harvard.edu/glensdata/

\refer Lane B. F., Creech-Eakman M., Nordgren T. E., 2002, ApJ, 573, 330 (astro-ph/0203060)

\refer Mould, J., et al., 2000, ApJ 529, 786 (astro-ph/9909260)

\refer Price et al., 2003, ApJ, 584, 931 (astro-ph/0207187)

\refer Weiler K.W., Sramek R.A., 1988, ARAA, 26, 295

\refer Witt H.J., Mao S., Keeton C.R., 2000, ApJ, 544, 98 (astro-ph/0004069)

\endrefer           
\end{document}